\documentclass[10pt,letterpaper]{article}
\usepackage[top=0.85in,left=2.75in,footskip=0.75in]{geometry}

\usepackage{tabularx}

\usepackage{amsmath,amssymb}

\usepackage{changepage}

\usepackage{textcomp,marvosym}

\usepackage{cite}

\usepackage{nameref,hyperref}

\usepackage[right]{lineno}

\usepackage[nopatch=eqnum]{microtype}
\DisableLigatures[f]{encoding = *, family = * }

\usepackage[table]{xcolor}

\usepackage{array}
\usepackage{tikz}
\definecolor{accent}{HTML}{B7222D}

\newcolumntype{+}{!{\vrule width 2pt}}

\newlength\savedwidth

\raggedright
\usepackage[aboveskip=1pt,labelfont=bf,labelsep=period,justification=raggedright,singlelinecheck=off]{caption}

\makeatletter
\renewcommand{\@biblabel}[1]{\quad#1.}
\makeatother

\usepackage{lastpage,fancyhdr,graphicx}
\usepackage{epstopdf}
\fancyheadoffset[L]{2.25in}
\fancyfootoffset[L]{2.25in}
\begin{document}
\vspace*{0.2in}

\begin{flushleft}
{\Large
\textbf{Recurrent losers: A network approach to cover-bidding patterns in public procurement} 
} \newline
\\
Isabela Villamil\textsuperscript{1*}, János Kertész\textsuperscript{1}
\\
\bigskip \textbf{1} Department of Network and Data Science, Central European University, Vienna, Austria
\\
\bigskip

%
%





* villamil\_isabela@phd.ceu.edu

\end{flushleft}


%
\begin{abstract}
We develop and apply a network-based method to examine patterns consistent with cover bidding within procurement markets, using only bidder participation and winner identities. A statistical test identifies relationships in which firms win unusually often when particular competitors also bid. These relationships form a directed graph with links from recurrent losers to recurrent winners. We use maximum $k$-core to classify possible organizational forms of cover bidding and propose tender exposure---the share of contested tenders involving a reported relationship---as a market-level indicator of potential cover-bidding risk.
Applied to 200,065 contracts procured by district engineering offices of the Philippine Department of Public Works and Highways (DPWH) from 2017 to 2025, covering three congressional terms, the method identifies 8,728 directed relationships across 407 of 564 markets. The $k$-core classes differ in network breadth, relationship depth, and bidder roles. In simpler networks, each recurrent loser is usually tied to a single winner, while highly interconnected networks involve multiple recurrent winners and broader groups of bidders that rarely win. Some of the deepest relationships and highest levels of tender exposure occur in intermediate networks, showing how relationships can reach much of a market without extensive interconnectedness.
Offices named in official investigations more often contain highly interconnected networks than other offices. Two of the three offices referred for bid rigging
are among the twelve in our ranking with the highest exposure in 2023--2025.
\end{abstract}

\clearpage
\newgeometry{top=0.85in,left=1in,right=1in,footskip=0.75in}

\section{Introduction}
Governments worldwide spend an estimated US$\$$13 trillion a year on public procurement (around 15 percent of global gross domestic product), making it one of the world's largest markets \cite{worldbank2021procurement}. Competition for public contracts affects not only the cost of government provision but also the quality and availability of public services. Bid rigging undermines this competition by allowing firms to coordinate prices and contract awards rather than competing independently. The resulting overcharges and inefficiencies divert public resources from other priorities, including infrastructure, education, and healthcare. Losses are difficult to measure, but estimates cited by the European Commission suggest that collusion can increase procurement costs by up to 60 percent relative to normal market conditions~\cite{ec2021collusion}.

One form of bid rigging that is particularly difficult to detect is cover bidding, in which firms submit deliberately losing offers to support a designated winner. Unlike bid suppression, cover bidding preserves the appearance of competition: several firms participate, but their bids are not independent. Because unsuccessful bids are also a normal outcome of competition (in fact, a large number of bidders is often considered a sign of healthy competition \cite{dpwhmanual2016v1}), individual tenders provide little evidence of such arrangements. However, across repeated tenders, systematic patterns may emerge. A firm that loses to some rivals and wins against others is unremarkable, but a lasting imbalance, in which one firm consistently loses to a particular competitor, makes their relationship suspicious.

The Philippines provides a particularly relevant setting for studying these patterns. Bid rigging in public works procurement has been documented for decades. A World Bank investigation of road contracts tendered by the Department of Public Works and Highways (DPWH) found bid patterns indicating collusion among contractors, and witnesses described prearranged winners and compensated losing bidders \cite{worldbank2010nrimp}. More recently, investigations beginning in 2025 into flood-control projects have led to bid-rigging referrals to the competition authority \cite{pcc2025inquiry,pcc2025doj}, and to allegations of graft and corruption involving officials, legislators, and politically connected contractors \cite{sandiganbayan2026cases,pcij2025flood}. These cases underscore longstanding concerns about competition and integrity in Philippine infrastructure procurement \cite{worldbank2008cpar}, and the need for systematic methods to examine bidding relationships across procurement markets.

We develop and apply a network-based approach to investigate possible organizational forms of cover bidding and to measure how extensively potential cover-bidding patterns occur within procurement markets. Using only the identities of participating bidders and the winner of each tender, we identify statistically unusual relationships between recurrent winners and recurrent losers and represent them as directed networks. We classify these networks by their maximum $k$-core and examine their breadth, relationship depth, and bidder roles to characterize possible arrangements for allocating contracts and supporting bids.

We propose \emph{tender exposure}, the proportion of eligible contested tenders involving at least one reported winning--losing relationship, as a market-level indicator of potential cover-bidding risk. We apply the approach to public works contracts awarded by DPWH district engineering offices between 2017 and 2025, and use publicly documented investigations as independent points of reference.

The study makes two contributions. First, it provides a framework for characterizing possible organizational forms of cover bidding. Conventional co-bidding networks are projections of bipartite firm--tender networks, connecting firms that participate in the same tenders. While these capture which firms bid together, our directed relationships identify which firms repeatedly lose to which competitors, allowing us to connect observed patterns of contract allocation to questions of cartel organization. Second, it develops an interpretable market-level indicator of potential cover-bidding risk that can help procurement authorities identify markets warranting further investigation. Both rely on routinely collected procurement records, making the approach usable even when the values of losing bids are unavailable.

As the literature has long recognized, competitive and collusive bidding can produce similar patterns of winning and losing. Persistent cost advantages or differences in available capacity may explain repeated success against the same competitors without coordination \cite{bajariye2003,kawai2023}. The method therefore cannot establish collusion from bidding outcomes alone. Its value lies in narrowing the field of investigation, where additional evidence on bids, ownership, and procurement decisions can help distinguish competition from coordination.

The remainder of the paper is organized as follows. Section 2 reviews the related literature, and Section 3 describes the Philippine procurement setting and investigations. Section 4 presents the data and methods. Section 5 reports the results, Section 6 discusses their implications, and Section 7 concludes.

\section{Related research}

The success of a cartel depends on four conditions: coordination, internal stability, external stability, and enforcement \cite{harrington2023}. Each shapes how collusion is organized and, in procurement markets, how contracts and supporting bids are distributed among participating firms. Coordination requires agreement on which firms will win and which will submit losing bids. At one extreme, a designated winner may rely on a small number of supporting bidders, sometimes through bilateral agreements \cite{oft2009construction,cat2011gmi}. Other arrangements involve several designated winners sharing supporting bidders or providing cover for one another \cite{conleydecarolis2016,comco2019}. The allocation of winning opportunities can thus be concentrated among a few firms or distributed across a larger group, requiring different degrees of coordination.

Reaching an agreement does not ensure that firms will continue to honor it. Internal stability depends on how participants are compensated for submitting losing bids. McAfee and McMillan distinguish strong bidding rings, which can make side payments and exclude entrants, from weak rings, which cannot make transfers \cite{mcafeemcmillan1992}. Without side payments, members must receive sufficient winning opportunities to make cooperation worthwhile \cite{pesendorfer2000,ishii2009,aoyagi2003}. Where transfers are possible, supporting bidders may instead receive payments or subcontracts and need not win contracts themselves \cite{pesendorfer2000,levensteinsuslow2006,oecd2022screening}. Common ownership offers another possibility: when nominally separate bidders share an owner, the benefits of winning accrue to the same interests regardless of which firm receives the contract \cite{conleydecarolis2016,vanklinken2010}. These mechanisms may leave different patterns in procurement records. Firms compensated through future awards may alternate between winning and losing, whereas those compensated outside the tender, or under common ownership, may repeatedly lose without receiving contracts.

External stability depends on whether a cartel can protect its agreement from competition by outsiders. A small group may be relatively easy to coordinate but vulnerable to firms that are not party to the arrangement. Expanding membership can reduce this threat, at the cost of accommodating more participants and potentially conflicting interests. Indeed, entry and difficulties in renegotiating agreements have historically posed greater threats to cartel survival than secret cheating \cite{levensteinsuslow2006}. Barriers to entry can therefore be important to the viability of relatively small arrangements, while broader participation may be necessary when competition from outsiders cannot otherwise be restricted. There is no reason to expect a single organizational size or allocation rule to resolve these competing pressures in every market.

The final condition, enforcement, concerns avoiding detection and penalties. Larger cartels may face greater difficulties concealing their activities, since more members and more extensive communication increase the chance of detection \cite{harrington2023}. Levenstein and Suslow observe that documented cartels with more members commonly relied on industry associations or government involvement to organize their activities \cite{levensteinsuslow2006}. In public procurement, officials may perform some of these organizational functions, designating winners, coordinating contract allocations, or restricting competition through tender requirements \cite{tanaka2016,lambertsonin2006}. Their involvement may also weaken the oversight intended to detect collusion. Hudon and Garz\'on describe procurement coalitions involving contractors, public officials, and political actors that circumvent monitoring and influence procurement decisions \cite{hudongarzon2016}. Official participation can therefore help sustain a bidding arrangement by facilitating coordination, protecting it from outside competition, and reducing the risk of detection.

The OECD distinguishes four bid-rigging strategies: cover bidding, in which firms submit bids designed to lose; bid suppression, in which firms refrain from bidding or withdraw their bids; bid rotation, in which members take turns winning; and market allocation, in which they divide customers or territories. These strategies may be used separately or in combination \cite{oecd2025}. Each leaves different traces in procurement records. Bid suppression and market allocation can reduce encounters between colluding firms, while rotation may distribute winning opportunities among members \cite{oecd2025,mcafeemcmillan1992,imhof2018}. Screens for these strategies examine unusually low bidder participation, patterns of firms avoiding one another's tenders, and systematic alternation in contract awards \cite{oecd2022screening}. Cover bidding, however, presents a particular detection problem because it is designed to give the appearance of genuine competition: firms continue to participate in tenders, but with no intention to win \cite{oecd2025}.

Distinguishing an imbalance caused by competition from one caused by cover bidding is a major scientific and practical challenge \cite{bajariye2003,harrington2008}. Most empirical screens for cover bidding examine the amounts of unsuccessful bids. Some examine the spacing and dispersion of bids \cite{imhof2018}, the absence of losing bids close to the winner \cite{chassang2022}, the persistence of designated winners across rebidding rounds \cite{kawai2022}, and the clustering of bids above an isolated winning bid \cite{clark2025}. More elaborate approaches test whether bids satisfy the conditional independence and exchangeability conditions implied by competitive bidding \cite{bajariye2003}, or test for coordinated entry and bidding decisions \cite{conleydecarolis2016}. These techniques offer different ways to distinguish coordinated from competitive bidding, but require information about unsuccessful bids or features of the auction process that are not always available in procurement records.

When unsuccessful bid amounts are unavailable, screens can instead examine bidder participation and contract awards. Competition authorities have developed indicators of firms that win unusually often or repeatedly participate without winning. The OECD documents screens used in Brazil, Korea, and Denmark that compare firms' winning frequencies or success against competing bidders \cite{oecd2022screening}. Such indicators may identify unusually successful or unsuccessful firms, but do not necessarily reveal whether their outcomes depend on the particular competitors they encounter. A firm may win frequently because it is more competitive than its rivals, while a firm with an otherwise unremarkable winning rate may consistently lose to one particular competitor.

Repeated encounters between bidders provide another source of evidence about possible coordination. Firms participating in collusive arrangements may bid together more frequently than would be expected under competition \cite{feinstein1985}. Network approaches extend this reasoning. Wachs and Kert\'esz~\cite{wachs2019} identify high-risk groups of firms in the co-bidding network: groups whose interactions are both coherent, that is, consistent and intense, and exclusive, that is, separated from the rest of the market. Network position may also be informative. In municipal construction tenders in Laval, Quebec, Reeves-Latour and Morselli~\cite{reeveslatour2017} find that firms suspected of bid rigging were perennial participants in the core of the co-bidding network. Using cartel members identified in a criminal investigation of Italian roadwork tenders, Carbone et al.~\cite{carbone2024} find that firms embedded in deeper $k$-cores of the co-bidding network are more likely to belong to a cartel. In buyer--supplier networks, Wachs et al.~\cite{wachs2021} find that single bidding, a corruption-risk indicator, is more common among the most central buyers and suppliers in some European countries and in the periphery in others. Medina-Hern\'andez et al.~\cite{medinahernandez2026} show that network features, particularly coreness and eigenvector centrality, improve the prediction of fraud and corruption risk in Mexican procurement. These approaches reveal patterns of participation and contracting.

Winning and losing records add direction to these relationships. Repeated auctions may reveal reciprocal exchanges of winning opportunities, persistent dominance by particular firms, or systematic differences in success against specific competitors \cite{ishii2009,learya2022,lenguyen2022}. More recent pairwise tests exploit bidding behavior to distinguish potentially coordinated relationships from competitive differences between firms. In multistage auctions, for example, bidders who repeatedly refrain from undercutting particular competitors despite opportunities to do so are substantially more likely to be associated with sanctioned collusion \cite{baranek2025}. Such evidence is particularly informative where detailed bidding histories are available.

\section{Investigations into DPWH procurement}\label{sec:institutional}

The Department of Public Works and Highways (DPWH) of the Philippines procures roads, flood-control works, and other infrastructure through central, regional, and district offices. Each district engineering office (DEO) administers procurement within its jurisdiction through a Bids and Awards Committee. Project funding is determined through the national budget, in which members of Congress participate through the appropriations process \cite{dbm2016budgetcycle}.

Investigations into Philippine flood-control projects beginning in 2025 raised allegations involving construction firms, legislators, and DPWH officials. These concerned alleged irregularities in the allocation of project funding, the selection of contractors, and the conduct of bidding \cite{pcc2025inquiry,gma2025referrals,sandiganbayan2026cases}.

Two groups of contractors received particular attention. The first involved firms associated with the family of Elizaldy ``Zaldy'' Co, an Ako Bicol party-list representative who chaired the House Committee on Appropriations from July 2022 to January 2025 \cite{porcalla2025cochair}. Co co-founded Sunwest with his brother Christopher, who also established Hi-Tone Construction \& Development Corporation~\cite{inquirer2025co,pcij2025flood}. Their sister Farida Co, Albay's vice-governor, was identified as the owner of FS Co Builders and Supply. Although the brothers formally divested their interests upon entering Congress, their founding and family connections remained subjects of scrutiny \cite{pcij2025flood}. In our district-office sample, the three firms, including their joint ventures, received contracts with recorded awards totaling approximately PHP~86.3 billion (Philippine pesos) during 2017--2025.

The second group comprised companies associated with Pacifico ``Curlee'' and Cezarah ``Sarah'' Discaya. Corporate records examined by the Philippine Center for Investigative Journalism {connected nine firms to the family} through {ownership or management} \cite{pcij2025flood}. Senate hearings in September 2025 examined the participation of Discaya-owned companies as separate bidders for the same projects \cite{senate2025discaya}.

Testimony also described alleged connections between contractors' activities and decisions by legislators and procurement officials. Former DPWH Undersecretary Roberto Bernardo and former Bulacan First District Engineer Henry Alcantara implicated legislators and engineering personnel in arrangements involving project funding and payments \cite{bacelonia2025alcantara,casilao2025bernardo}. Numerous cases and investigations have since been initiated, concerning alleged conduct extending from budget allocation to contract award and involving actors both inside and outside the procuring authority.

\section{Materials and methods}

\subsection{Data and market definition}

We obtained contract-award and bidding records for 2017--2025 from the DPWH Transparency Portal \cite{dpwhportal}. Restricting the analysis to DEOs leaves 200,065 awarded contracts and 615,830 bidder appearances. Regional and central offices are excluded because they operate at different geographic and administrative scales and procure systematically larger or more specialized projects.

Procurement years are assigned to the Congress that enacted the corresponding national budget: 2017--2019 to the 17th Congress, 2020--2022 to the 18th, and 2023--2025 to the 19th. Each procurement market comprises the tenders awarded by one district engineering office during one Congress. The office identifies the procuring authority, while congressional terms align the observation periods with the national budget cycle and provide common three-year windows. Political turnover can also reorganize contracting relationships \cite{fazekaswachs2020}; applying the method separately within these markets allows bidder relationships to vary across offices and periods. Repeated participation is geographically localized (S1 Fig), supporting the use of office-level markets rather than a single national pool of bidders. Combining each DEO with each Congress produces 564 procurement markets. The data were extracted on 13 December 2025. The final observation window is incomplete: the snapshot contains awards dated through 16 October 2025, and awards fell sharply after the flood-control investigations began in August 2025. In August and September 2025, 354 contested contracts were awarded, compared with 1,294 in the same months of 2024.

We use \emph{contract} when describing the awarded project and \emph{tender} when describing its bidding process. Each contract corresponds to one tender. The unit of observation is a \emph{bidding actor}: either a firm bidding alone or a joint venture. Each joint venture is treated as one bidding actor, identified by its combination of partners. The same combination is treated as the same actor across tenders. A firm bidding independently and a joint venture involving that firm are therefore treated as distinct bidding actors. Where the records do not identify a joint venture, bidders are treated as separate. After reconstruction, the data contain 607,150 actor--tender records, 10,592 distinct actors and exactly one winning actor on every contract. Firm-level comparisons credit a joint venture's relationships and tenders to each recorded member.

Of the 200,065 contracts, 162,884 have at least two reconstructed bidding actors and are termed \emph{contested}. Single-bidder contracts cannot contain a loser--winner relationship and are used only to describe the procurement system, such as the share of contracts with a single bidder. Of the contested contracts, 197 fall in seven markets where no pair meets the minimum co-bidding requirement for testing. Because no relationship can be reported in these markets, they are excluded from the tender-exposure denominator, leaving 162,687 eligible contracts.

\subsection{Directed network construction}

\begin{figure}[!ht]
    \centering
    \includegraphics[width=\linewidth]{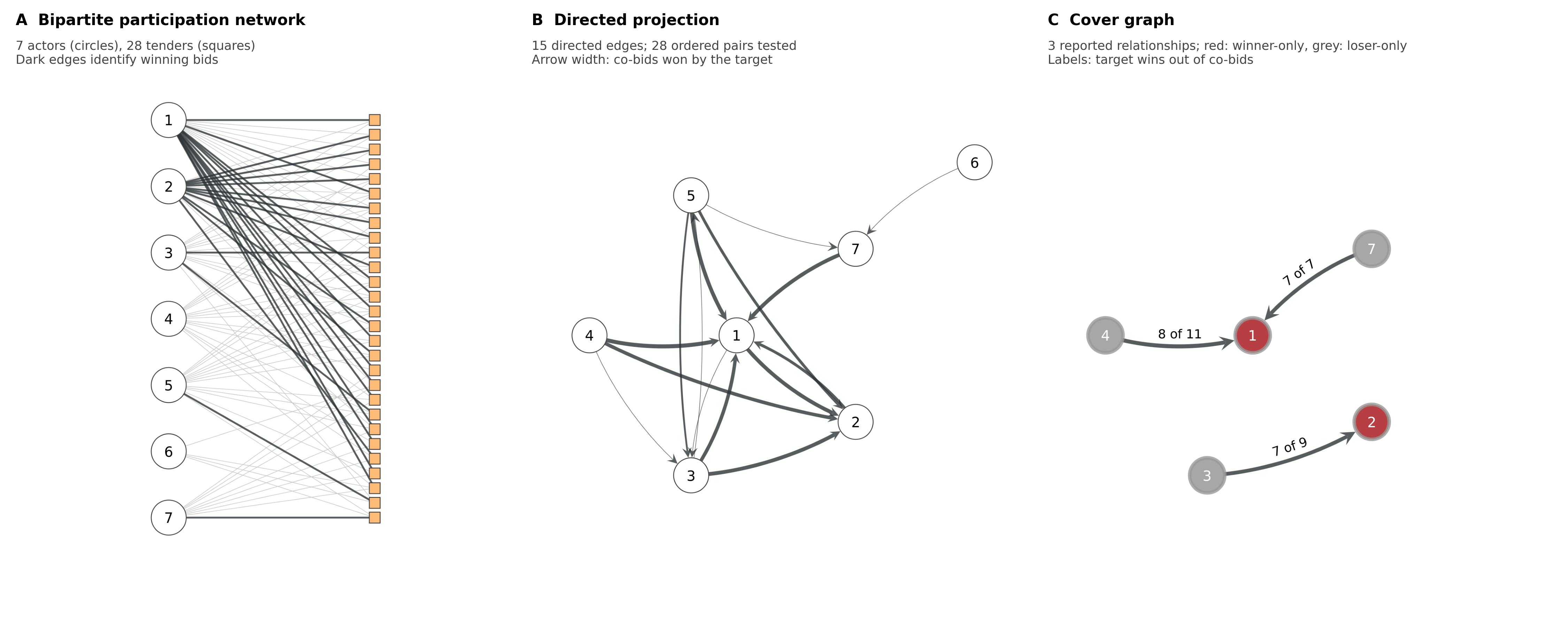}
    \caption{\textbf{From bidder participation to the cover graph.} A constructed market contains seven bidding actors and 28 tenders. \textbf{A}, the bipartite participation network $G_{\mathrm{B}}$; dark edges indicate winning bids. \textbf{B}, the directed projection $G_{\mathrm{P}}$; arrow widths indicate the number of co-bids won by the target ($N_{AB}$). \textbf{C}, the cover graph $G_{\mathrm{C}}$, retaining edges from $G_{\mathrm{P}}$ in which the target wins significantly more often than expected under equal winning probabilities (within-market Benjamini--Hochberg correction, target false discovery rate 5\%, $N_{AB}\geq3$). In Panel C, edge labels show the target's wins ($N_{AB}$) out of the pair's co-bids ($C_{AB}$); red nodes are winner-only and gray nodes are loser-only.}
    \label{fig:pipeline}
\end{figure}

{Within each DEO $\times$ Congress market, we apply a screen to identify pairs of bidders in which one wins unusually often when both participate. Fig~\ref{fig:pipeline} illustrates the procedure.

Panel A represents the observed bidding records as a bipartite network $G_{\mathrm{B}}$, with one set of nodes representing bidders and the other representing tenders. An edge connects a bidder to a tender they entered, with winning bids distinguished from losing bids.

Panel B shows $G_{\mathrm{P}}$, the directed projection of $G_{\mathrm{B}}$ onto the set of bidders. For two bidders $A$ and $B$, let $T_{AB}$ denote the set of tenders they entered together and $C_{AB}=|T_{AB}|$ the number of such tenders. Let $N_{AB}$ be the number won by $B$, and therefore lost by $A$. The projection includes an edge $A\to B$ for each ordered pair with $C_{AB}\geq3$ and $N_{AB}>0$. Its weight is $N_{AB}$.

Panel C shows the \emph{cover graph} $G_{\mathrm{C}}$, a subgraph of $G_{\mathrm{P}}$ containing the relationships in which $B$ wins unusually often against $A$, which we consider a signal of cover bidding. These relationships satisfy the statistical test and reporting rule described below.

To identify these edges, we compare the observed winning outcomes against a null model that preserves the participants in each tender but assigns them equal probabilities of winning. Each of the $n_t$ bidders in tender $t$ has a probability $1/n_t$ of winning, independently across tenders. The model conditions on observed participation but does not account for differences in firms' costs or capabilities. (This limitation will be discussed in Section 6.)

Under the null model, the number of tenders won by $B$ against $A$ is the random variable
\[
  X_{AB}=\sum_{t\in T_{AB}}X_t,
  \qquad X_t\overset{\mathrm{ind}}{\sim}
  \operatorname{Bernoulli}\!\left(\frac{1}{n_t}\right),
\]
where $X_t=1$ if $B$ wins tender $t$ under the null model, and zero otherwise. Because $n_t$ can differ across tenders, the win count $X_{AB}$ follows a Poisson--binomial rather than an ordinary binomial distribution. The one-sided p-value is the exact probability of obtaining at least the observed $N_{AB}$ wins:
\[
  p_{AB}=\Pr\!\left(X_{AB}\geq N_{AB}\right).
\]

We apply the Benjamini--Hochberg procedure \cite{benjaminihochberg1995} to these p-values separately within each market at a target false discovery rate of 5\%. The correction includes both directions of every pair with $C_{AB}\geq3$, including those with no observed wins. An edge $A\to B$ is retained in $G_{\mathrm{C}}$ if its null hypothesis is rejected after correction and $N_{AB}\geq3$. This minimum-win requirement is applied after correction. Each retained edge represents a \emph{reported relationship}, directed from $A$ to $B$ and weighted by $N_{AB}$.
}

\subsection{Network measures and comparisons}\label{sec:measures}

We distinguish bidding actors by their positions in the cover graph. Winner-only actors have incoming relationships but no outgoing relationships (``sinks''), loser-only actors have outgoing relationships but no incoming relationships (``sources''), and mixed-role actors have both. These classifications describe their roles in the reported relationships, rather than their overall winning or losing records.

{To characterize the cover graph $G_{\mathrm{C}}$, we distinguish two dimensions: breadth and depth. \emph{Graph breadth} is the number of nodes in $G_{\mathrm{C}}$, representing actors involved in at least one reported relationship. \emph{Relationship depth} is the edge weight $N_{AB}$, measuring the number of tenders won by $B$ against $A$. Breadth thus captures how many actors are linked, while depth captures the number of repeated winning--losing outcomes associated with each relationship.}

We use two measures to characterize potential cover-bidding arrangements at the market level. First, we use maximum $k$-core to classify how relationships are organized. A $k$-core is a maximal subgraph in which every actor has at least $k$ neighbors within it \cite{seidman1983}. For this calculation, we remove direction and weight from $G_{\mathrm{C}}$, treating relationships reported in both directions as a single undirected edge. The market's maximum $k$-core, its degeneracy, separates three configurations. It equals 1 if and only if the graph contains no cycle \cite{lickwhite1970}, so that relationships form only
trees, e.g., pairs, stars, and chains; a 2-core exists once supporting relationships close a loop, each member of the core being tied to at least two others; and in a $k\geq3$ core every member has at least three counterparts within it. We classify markets
as having no reported relationships, maximum $k=1$, maximum $k=2$, or maximum $k\geq3$. The class describes the market's most interconnected part rather than all members in the cover graph---in markets with maximum $k\geq3$, many linked actors also lie outside the highest core.

Second, we calculate \emph{tender exposure}, an indicator of potential cover-bidding risk that measures how extensively reported winning--losing relationships recur across a market's contested tenders. {For market $m$, let $\mathcal{E}_m$ be the set of edges of its cover graph $G_{\mathrm{C}}$ (Fig~\ref{fig:pipeline}C), so that $(A,B)\in\mathcal{E}_m$ is a reported relationship $A\to B$.} Exposure is defined as

\[
\mathrm{Exposure}_m=\frac{1}{|T_m|}\sum_{t\in T_m}I_t,
\]

where

\[
I_t=
\begin{cases}
1, & \text{if }(A,w_t)\in\mathcal{E}_m\text{ for some }A\in P_t,\\
0, & \text{otherwise}.
\end{cases}
\]

Here, $T_m$ is the set of eligible contested tenders in market $m$, $w_t$ is the winner of tender $t$, and $P_t$ is its set of participating bidders. {Exposure is thus the proportion of eligible contested tenders in which at least one losing bidder has a reported relationship directed toward the winner.} Each tender is counted once, regardless of the number of such relationships it contains.

We also examine how contract budgets and proximity to delegated approval thresholds differ across markets with different cover-graph structures and levels of exposure. {DPWH department orders set limits on the value of projects that district engineering offices may approve, procure, and implement; projects exceeding these limits fall under higher levels of authority. From October 2016, the applicable district-level limits were PHP 50 million or PHP 100 million, depending on the office \cite{dpwhdo208}. The higher limit increased to PHP 150 million in late 2022 \cite{dpwhdo195,dpwhdo250}. Keeping a project's budget within the applicable limit can therefore allow it to remain under district-level authority. DPWH regulations also prohibit splitting certain infrastructure projects into phases merely to fit an office's delegated authority \cite{dpwhdo250}.} We compare the distributions of approved budgets for the contract (ABCs) and the shares of contracts whose ABCs fall within 5\% below these thresholds, including the threshold itself.

We examine PHP 50 million and PHP 100 million in the 17th and 18th Congresses, adding PHP 150 million in the 19th. {We retain the earlier thresholds because project budgets may continue to reflect previously applicable limits, particularly when prepared before regulatory changes took effect. Since approval limits varied across offices and over time, we examine concentrations around these thresholds without assuming that each applied to every contract.}

Finally, we compare the cover-graph structure and tender exposure of ten district offices linked to DPWH bid-rigging referrals and to charges before the Sandiganbayan (the Philippine anti-graft court) with those of other procurement markets. We also examine the reported relationships involving Discaya-linked and Co-linked firms to characterize their bidding behavior and assess whether they exhibit anomalous patterns under the network-based approach.

Analyses were carried out in Python 3.10, using graph-tool 2.98 \cite{peixoto2014graphtool} for graph measures and SciPy 1.15 for rank correlations.

\section{Results}

\subsection{Reported relationships}

The screen identifies 8,728 directed relationships in 407 of the 564 DEO $\times$ Congress markets, with the breakdown by Congress reported in Table~\ref{tab:national}. These relationships range from a few co-bids to hundreds within the same procurement office and Congress.

\begin{table}[!ht]
\centering
\caption{\textbf{Procurement records and screening results, by Congress.} Each market is one district engineering office during one Congress. Procurement years are assigned to the Congress that enacted the corresponding national budget: 2017--2019 to the 17th Congress, 2020--2022 to the 18th, and 2023--2025 to the 19th. A bidding actor is either a firm or a joint venture, with each joint venture treated as one actor. Because the same actor may bid in more than one Congress, the All column reports 10,592 unique actors rather than the sum of the three Congress-specific counts.}
\small
\begin{tabular}{lrrrr}
\hline
 & 17th & 18th & 19th & All \\
\hline
\multicolumn{5}{l}{\textit{Procurement records}} \\
Awarded contracts & 71,667 & 62,298 & 66,100 & 200,065 \\
Contracts with at least two bidding actors & 56,687 & 51,652 & 54,545 & 162,884 \\
Bidding actors & 5,412 & 6,068 & 6,846 & 10,592 \\
Markets (DEO $\times$ Congress) & 183 & 185 & 196 & 564 \\
\hline
\multicolumn{5}{l}{\textit{Screening results}} \\
Tested relationships, $A\to B$ & 90,784 & 74,692 & 79,892 & 245,368 \\
Reported relationships & 3,018 & 2,613 & 3,097 & 8,728 \\
Markets with at least one reported relationship & 130 & 129 & 148 & 407 \\
\hline
\end{tabular}
\label{tab:national}
\end{table}

In 84\% of the reported relationships, $A$ wins none of the tenders entered by both $A$ and $B$. These pairs co-bid an average of 12 times, although some meet far more often. In Isabela 1st DEO during the 19th Congress, two firms entered 171 of the same tenders. One won 164, while the other won none.

At the other extreme, some relationships are reported after only three co-bids, all won by $B$. These tenders attract about 15 bidders on average, making three wins by the same firm unlikely under our null model. The screen therefore identifies relationships supported by very different lengths of bidding history, from a few highly asymmetric encounters to patterns repeated across hundreds of tenders.

\begin{figure}[!ht]
    \centering
    \includegraphics[width=\linewidth]{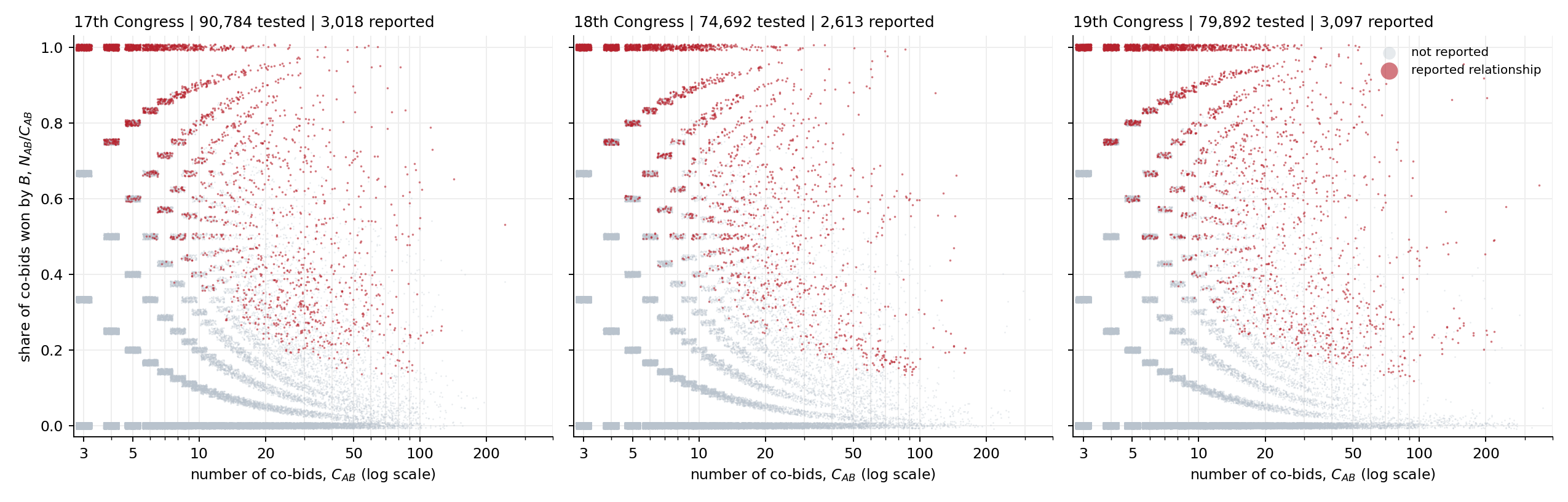}
    \caption{\textbf{Reported relationships by co-bidding frequency and winning share.} Each point represents one tested direction, $A\to B$, within a DEO $\times$ Congress market. The horizontal axis shows how many tenders $A$ and $B$ entered together; the vertical axis shows the proportion won by $B$. Red points are reported relationships; gray points are pairs not reported by the screen. When actors co-bid only a few times, reported relationships generally require $B$ to win nearly every tender. With more co-bids, relationships can be reported at lower winning shares. Of the 245,368 tested directions, 8,728 (3.6\%) are reported. Points are slightly displaced to reveal overlapping observations.}
    \label{fig:detection}
\end{figure}

Fig~\ref{fig:detection} shows how the reported relationships are distributed across the number of co-bids and the proportion won by $B$. Among pairs that meet only a few times, reported relationships are almost entirely one-sided. With more shared tenders, relationships can be reported even when $B$ does not win every encounter.

\subsection{Bidder roles and cover-graph structure}\label{sec:structure}

Of the 1,979 actors involved in reported relationships, 855 appear only as recurrent losers and 624 only as recurrent winners. The remaining 500 occupy both roles against different counterparts; 262 do so within the same office and Congress. In Abra during the 19th Congress, firm $L$ won 241 of 376 tenders entered with firm $J$, but lost 172 of 198 tenders entered with firm $Q$ (S2 Fig). Opposing relationships can also occur within the same pair. In Quezon 4th DEO during the 18th Congress, firms $B$ and $P$ entered 145 tenders together and each won almost half, with both $P\to B$ and $B\to P$ reported (S3 Fig). Winning and losing roles thus depend on the counterpart rather than representing fixed characteristics of a firm.

The cover graphs differ in how these relationships are organized. Of the 407 markets containing reported relationships, 233 have maximum $k=1$, 115 have $k=2$, and 59 have $k\geq3$. Graphs with maximum $k=1$ consist of pairs, chains, or stars. In these, 89\% of recurrent losers are connected to a single winner. Some are nevertheless extensive: {in Romblon in the 19th Congress, one recurrent winner (the Co-linked firm Sunwest) has 14 recurrent losers (S1 File, p.~5)}. Markets with maximum $k=2$ contain at least one cycle of relationships. Their highest cores contain seven actors on average and commonly include recurrent losers connected to multiple winners. In markets with maximum $k\geq3$, each actor in the highest core has at least three connections within that core. These cores contain ten actors on average, with recurrent losers typically connected to three or four winners. The classes distinguish relatively simple configurations from more interconnected groups in which several recurrent winners have overlapping relationships with losers.

The winning records of loser-only actors provide further insight into their roles in these networks (Table~\ref{tab:loser_winrates}). In $k=1$ markets, these actors win an average of 12\% of their bids despite repeatedly losing to particular rivals. This suggests that their reported relationships do not necessarily reflect a general inability to compete. In $k\geq3$ markets, by contrast, 70\% of loser-only actors never win a contested tender in their focal market, and their mean win rate is only 3\%. They are more successful in other offices, with a mean win rate of 15\%, although this remains below the 23\% observed for loser-only actors in $k=1$ markets. Markets with $k=2$ fall between these profiles. These differences suggest variation in how bidder roles are distributed across structural classes: many recurrent losers in simpler networks remain capable of winning contracts elsewhere, while those in more interconnected networks may primarily serve supporting roles.

\begin{table}[!ht]
\centering
\caption{\textbf{Winning records of loser-only actors by cover-graph structural class.}
Loser-only actors appear exclusively as recurrent losers in a DEO $\times$ Congress market. Win rates are calculated over all contested tenders in 2017--2025 that a firm entered, separately for the market in which it occupies this role (focal market) and for other DEOs. The latter include only firms with bidding activity outside the focal office. The same firm may appear in multiple markets.}
\label{tab:loser_winrates}
\begin{tabular}{lccc}
\hline
Measure & $k=1$ & $k=2$ & $k\geq3$ \\
\hline
Average win rate in focal market (\%) & 12 & 8 & 3 \\
Average win rate in other offices (\%) & 23 & 20 & 15 \\
Proportion of loser-only actors with no wins in focal market (\%) & 37 & 51 & 70 \\
\hline
\end{tabular}
\end{table}

\begin{figure}[!ht]
    \centering
    \includegraphics[width=\linewidth]{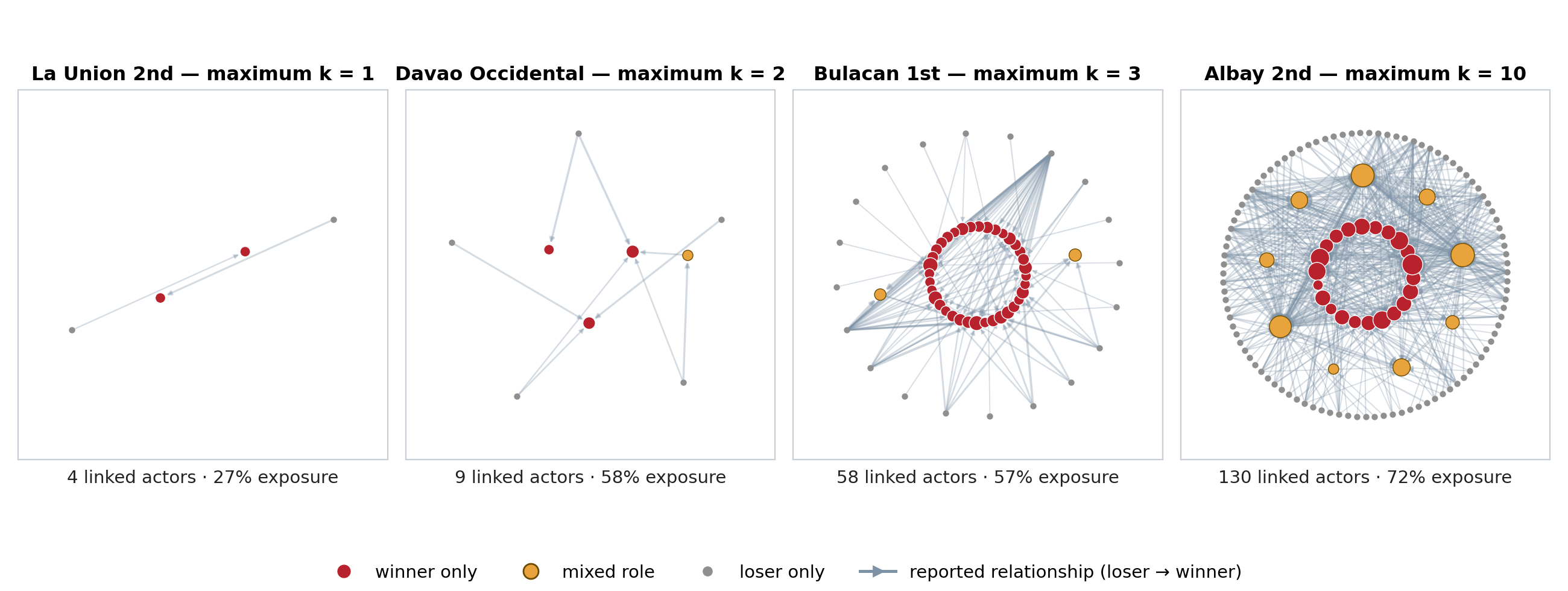}
    \caption{\textbf{Cover graphs with increasingly interconnected cores.} Four DEO markets from the 19th Congress, ordered by maximum $k$-core. Arrows point from recurrent loser to recurrent winner. Red nodes are winner-only, gray loser-only, and amber mixed-role. Thicker arrows represent more repeated losses. Labels below each panel give the number of linked actors and the share of contested tenders exposed to a reported relationship.}
    \label{fig:illustrated_four}
\end{figure}

Fig~\ref{fig:illustrated_four} illustrates four configurations of different structural classes (all 407 cover graphs are drawn in S1 File). {The offices were chosen for their links to the public cases: La Union 2nd, Davao Occidental, and Bulacan 1st were named in DPWH's 2025 bid-rigging referrals to the Philippine Competition Commission \cite{pcc2025inquiry,gma2025referrals}, and Albay 2nd is where the Co-linked firms have their largest presence, winning 45\% of the office's award value across the three Congresses.} La Union 2nd ($k=1$) consists of two separate pairs, while Davao Occidental ($k=2$) contains a three-actor core in which each actor is connected to the other two. Bulacan 1st ($k=3$) has several recurrent winners connected through overlapping losing relationships, including an actor that repeatedly loses to 30 winners. Albay 2nd ($k=10$) has the highest maximum core in the sample. The examples show how recurrent winning--losing relationships extend from separate bilateral configurations to highly interconnected groups, with variation in the depth of individual ties.

\subsection{Graph breadth, relationship depth, and tender exposure}

Graph breadth increases sharply across the structural classes. Markets with maximum $k=1$ have an average of 6.5 linked actors, compared with 48 in markets with $k\geq3$. Part of this difference reflects market size, since larger markets offer more opportunities for repeated encounters. However, even among markets of similar size, the proportion of actors in a tested pair that are linked varies considerably (S4 Fig). The largest bidder fields occur in a few highly interconnected markets. The ten markets with the highest average number of bidders per contested tender are all in Albay and Sorsogon, with average fields of 10 to 22 bidders. Nine of these markets have maximum $k\geq3$.

\begin{figure}[!ht]
    \centering
    \includegraphics[width=\linewidth]{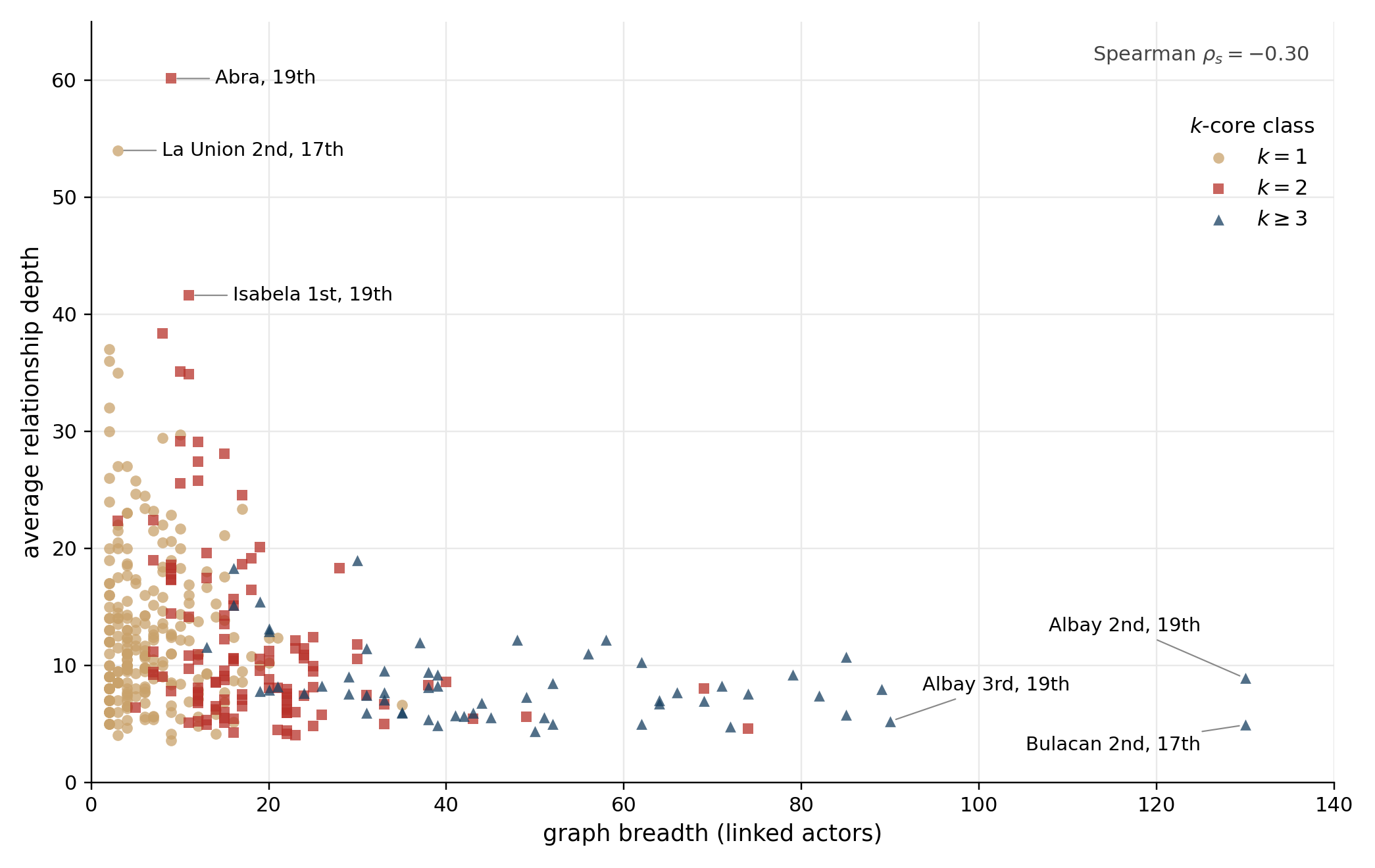}
    \caption{\textbf{Graph breadth and relationship depth.} Each point is a DEO $\times$ Congress market with at least one reported relationship. Graph breadth is the number of linked actors; average relationship depth is the mean number of repeated losses per reported relationship. Colors and symbols identify the maximum $k$-core classes. The three markets with the deepest average relationships and the three with the broadest graphs are labeled.}
    \label{fig:breadth_depth}
\end{figure}

The broader networks also tend to have shallower individual relationships. Average relationship depth falls from 12.8 repeated losses at $k=1$ to 12.3 at $k=2$ and 8.5 at $k\geq3$. Markets with deeper relationships generally have fewer linked actors (Spearman's $\rho_s=-0.30$), although considerable variation remains within each structural class (Fig~\ref{fig:breadth_depth}).

\clearpage

Across all markets, 22\% of eligible contested tenders are exposed to at least one reported winning--losing relationship. These tenders, worth about PHP 986 billion in 2017--2025, account for 28\% of the award value of eligible contested contracts because they tend to be larger (Section~\ref{sec:budgets}). Mean exposure increases with network interconnectedness, from 16\% in markets with maximum $k=1$ to 34\% at $k=2$ and 44\% at $k\geq3$.

\begin{figure}[!ht]
    \centering
    \includegraphics[width=0.82\linewidth]{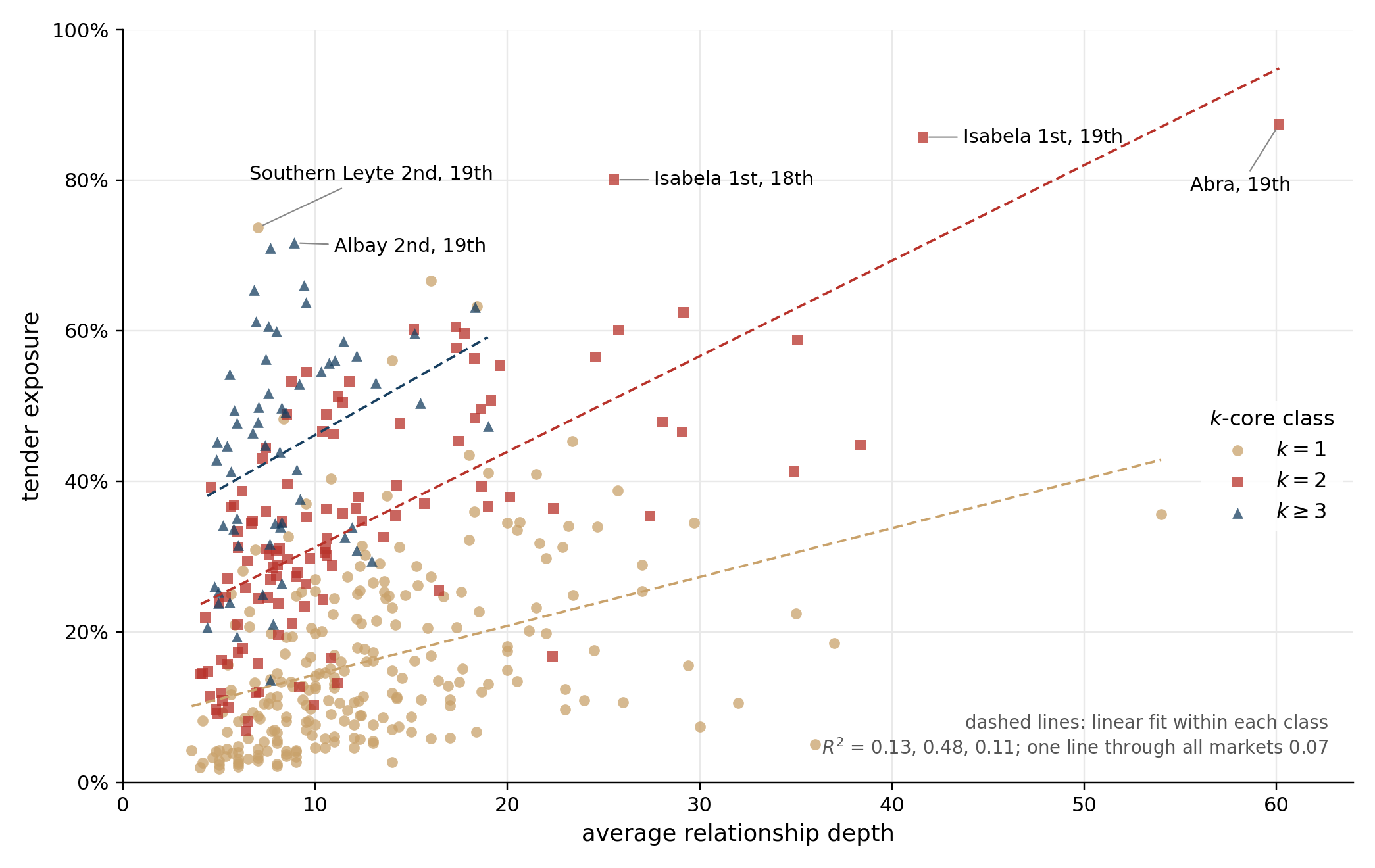}
    \caption{\textbf{Relationship depth and tender exposure.} Each point is a DEO $\times$ Congress market with at least one reported relationship. The horizontal axis shows the market's average relationship depth. The vertical axis shows the share of contested tenders on which a reported relationship connects the winner to a losing bidder. Colors and symbols identify the $k$-core classes; dashed lines are class-specific linear fits, with $R^2$ values given in the panel. The five markets with the highest exposure are labeled.}
    \label{fig:depth_exposure}
\end{figure}

Fig~\ref{fig:depth_exposure} shows how exposure varies with relationship depth across the structural classes. Within each class, markets with deeper relationships tend to have higher exposure. But exposure also differs substantially between classes at comparable depths. Among markets with an average relationship depth of 8--12 repeated losses, mean exposure is 13\% at $k=1$, 33\% at $k=2$, and 48\% at $k\geq3$.

The most exposed markets show how differently high exposure can arise. Abra and Isabela 1st each have exposure exceeding 85\%, arising from approximately a dozen deep relationships. Albay 2nd reaches 72\% exposure through 493 predominantly shallower relationships. Thus, a small number of highly persistent relationships and a much broader network of less persistent relationships can both account for substantial portions of a market's procurement activity. The markets that combine the highest exposure with the deepest relationships are mostly $k=2$ (Fig~\ref{fig:depth_exposure}). Some offices remain highly exposed across Congresses: Isabela 1st, Quirino, and Albay 1st are the only offices with exposure above 50\% in all three.

\begin{figure}[!ht]
    \centering
    \includegraphics[width=0.9\linewidth]{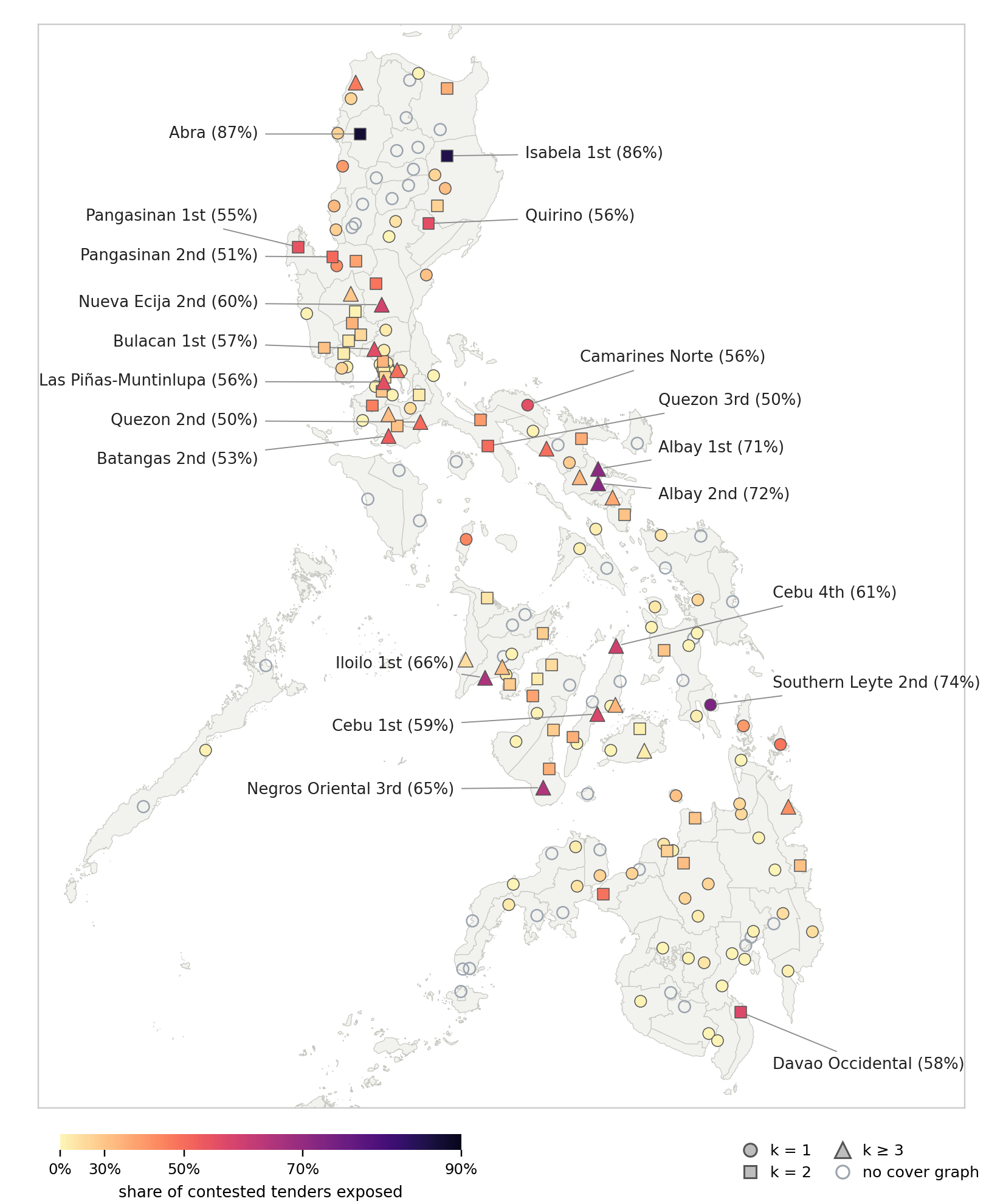}
    \caption{\textbf{Geographical distribution of tender exposure and maximum
    $k$-core across district engineering offices, 19th Congress.}
    Marker colors represent tender exposure and marker shapes indicate the
    maximum $k$-core class; hollow circles mark offices without a cover graph. Offices with exposure of 50\% or more are named, with their exposure in parentheses. Administrative boundaries from the UN OCHA common operational dataset for the Philippines (CC BY-IGO).}
    \label{fig:exposure_map}
\end{figure}

Fig~\ref{fig:exposure_map} maps tender exposure and maximum $k$-core across district engineering offices during the 19th Congress. Highly exposed offices occur across the three Philippine island groups (Luzon, Visayas, and Mindanao), but differ in network structure. Abra and Isabela 1st combine high exposure with intermediate-core networks, whereas Albay 1st and Albay 2nd exhibit highly interconnected networks.

\subsection{Contract budgets and approval ceilings}\label{sec:budgets}

\begin{figure}[!htbp]
    \centering
    \includegraphics[width=\linewidth]{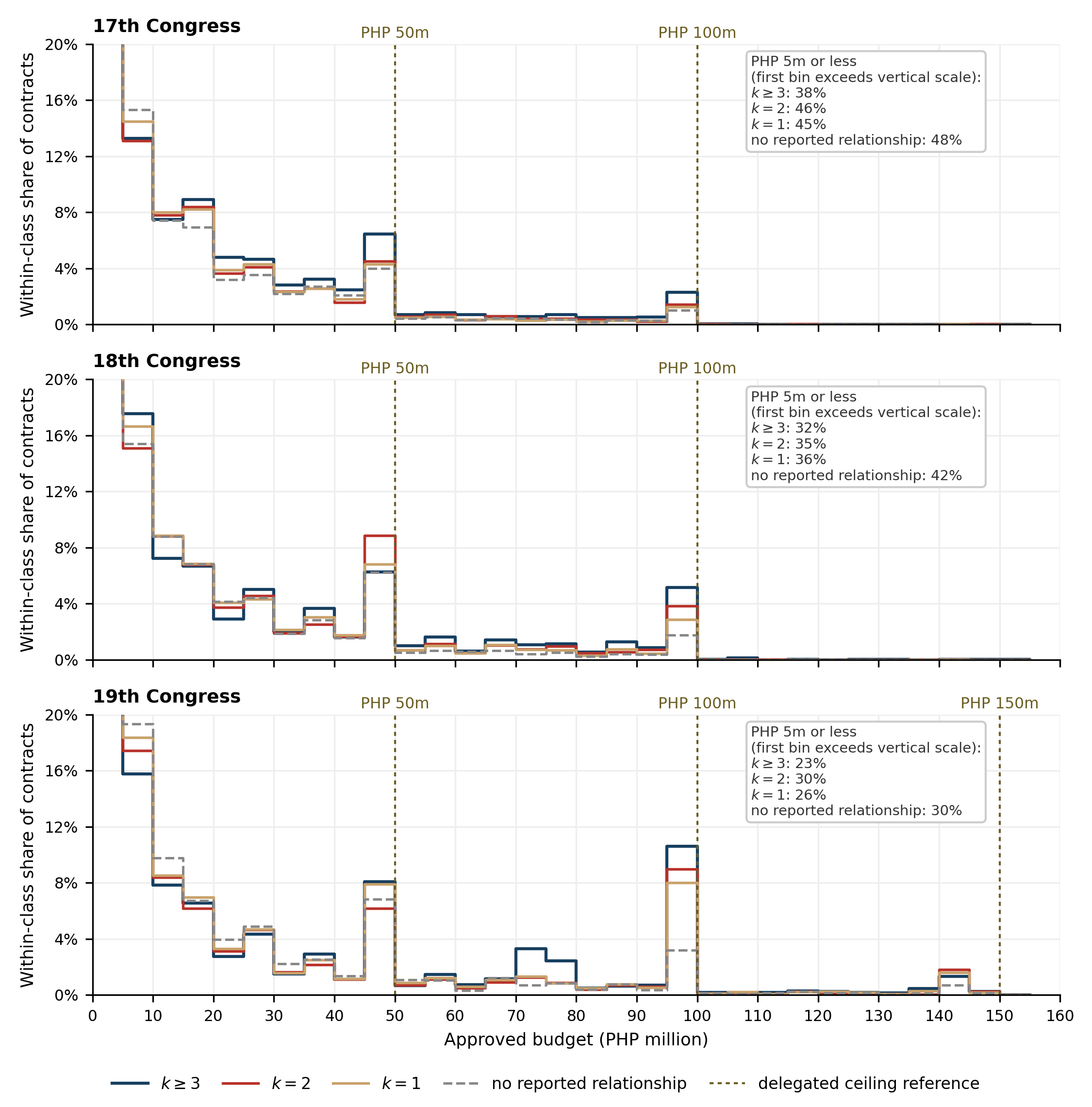}
    \caption{\textbf{Contract budgets by cover-graph class.} Each curve shows the distribution of approved budgets among contested contracts awarded in DEO~$\times$~Congress markets of the indicated class, using PHP 5 million bins; contracts, rather than markets, receive equal weight. Rows distinguish Congresses. Vertical dotted lines mark the delegated approval-ceiling reference amounts used in the analysis. The first bin exceeds the 20\% vertical range; the inset reports the share at PHP 5 million or less. Contracts above PHP 160 million are not displayed because of their rarity. Sample sizes and summary statistics are reported in S1 Table.}
    \label{fig:budgets_class}
\end{figure}

Markets with more interconnected cover graphs tend to award larger contracts and have a greater share of contracts budgeted immediately below delegated approval thresholds (Fig~\ref{fig:budgets_class}). During the 19th Congress, contracts in markets with maximum $k\geq3$ had a mean approved budget for the contract of PHP 36 million, compared with PHP 24 million in markets without reported relationships. The proportion budgeted within 5\% below the approval thresholds was also higher, at 20\% compared with 10\%. Both differences appear across all three Congresses.

The approval thresholds are particularly relevant because they determine the level of authority required to approve a contract. As discussed in Section~\ref{sec:measures}, DPWH implemented delegated approval limits of PHP 50 million and PHP 100 million (depending on the office) in October 2016, and introduced a PHP 150 million limit in late 2022 \cite{dpwhdo208,dpwhdo195,dpwhdo250}. The budget distributions change around these amounts. Before the PHP 150 million limit was introduced, only 2--20 contested contracts per year fell within 5\% below it. The number increased to 91 in 2023, 408 in 2024, and 404 in 2025 (S2 Table).

Not every concentration of contract budgets corresponds to an approval threshold. For example, the concentration between PHP 70 million and PHP 80 million in highly interconnected markets during the 19th Congress largely reflects 295 flood-control contracts awarded by Bulacan 1st DEO, with budgets clustered near PHP 72.4 million or PHP 77.2 million. In 2025, the Philippine Competition Commission referred a case involving this office to the Department of Justice after finding preliminary evidence that contractors had predetermined winning and losing bidders with the assistance of district officials \cite{pcc2025doj}. Berse et al. \cite{berse2025} also identify Bulacan 1st DEO as having the highest number of flood-control projects flagged for possible contract splitting or duplication.

Contract budgets and proximity to approval thresholds also vary with tender exposure. During the 19th Congress, 22\% of exposed contracts were budgeted immediately below an approval threshold, compared with 16\% of unexposed contracts in markets with cover graphs and 10\% of contracts in markets without them. Mean budgets followed the same pattern: PHP 38 million, PHP 30 million, and PHP 24 million, respectively (S5 Fig; S1 Table). Thus, the association with larger budgets and proximity to approval thresholds is strongest for contracts awarded to recurrent winners when a corresponding recurrent loser also submits a bid.

\clearpage
\subsection{Comparisons with documented cases}\label{sec:cases}

We examine the results for ten district offices named in publicly documented investigations and charges (Table~\ref{tab:official}). Three were referred by DPWH to the Philippine Competition Commission for suspected bid rigging \cite{pcc2025inquiry,gma2025referrals}. In three others, officials have been charged before the Sandiganbayan with graft and plunder over alleged kickbacks, and the remaining four awarded contracts named in conflict-of-interest charges against two legislators \cite{sandiganbayan2026cases}. These cases do not provide an exhaustive record of suspected misconduct, as other investigations may be ongoing or not publicly disclosed. Moreover, the allegations concerning an office may relate to only some of its procurement activities. Nevertheless, the documented cases provide independent points of reference for examining the relationships identified by the screen.

All ten DEOs in Table~\ref{tab:official} have reported winning--losing relationships in the 19th Congress. Four of the ten---Bulacan 1st, Iloilo 1st, Tarlac, and Antique---have maximum $k\geq3$, compared with approximately one in seven of the remaining offices with a cover graph. This difference persists when offices of similar procurement volume are compared.

Tender exposure varies considerably among the ten offices. Bulacan 1st, Davao Occidental, and Iloilo 1st are among the twelve most exposed offices nationally, with exposure between 57\% and 66\%. Five others fall below the 26\% mean among the other offices with cover graphs in the 19th Congress. Thus, although all ten offices have reported relationships, the proportion of their contested tenders involving these relationships differs substantially.

\begin{table}[!ht]
\centering
\caption{\textbf{Cover-graph characteristics of district offices named in
official actions.} All values refer to the 19th Congress. Maximum relationship
depth is the largest number of repeated losses in a reported relationship.
Exposure is the share of contested tenders containing a reported relationship.
The final row gives means across the remaining 138 offices with a cover graph.}
\small
\resizebox{\linewidth}{!}{%
\begin{tabular}{llrrrrr}
\hline
Documented official action & Office & Contested tenders & Max. $k$ & Linked actors & Maximum depth & Exposure \\
\hline
Bid-rigging referral & Bulacan 1st      & 1,559 & 3 & 58 & 62 & 57\% \\
                     & Davao Occidental &   194 & 2 &  9 & 26 & 58\% \\
                     & La Union 2nd     &    98 & 1 &  4 & 18 & 27\% \\
Kickback charges     & South Manila     &   325 & 2 & 22 &  8 & 15\% \\
                     & Quezon City 1st  &   594 & 1 &  6 & 21 &  9\% \\
                     & North Manila     &   566 & 1 &  4 & 13 &  5\% \\
Conflict-of-interest charges & Iloilo 1st & 312 & 4 & 38 & 53 & 66\% \\
                     & Tarlac           &   914 & 3 & 20 & 52 & 29\% \\
                     & Antique          &   162 & 3 & 19 & 18 & 21\% \\
                     & Iloilo 2nd       &   461 & 1 &  9 & 29 &  7\% \\
\hline
Remaining offices with a cover graph (mean) &                    &   309 & 1.7 & 15.0 & 26.7 & 26\% \\
\hline
\end{tabular}%
}
\normalsize
\label{tab:official}
\end{table}

{We next examine the relationships involving the nine Discaya-linked and three Co-linked firms introduced in Section~\ref{sec:institutional}. Together, these twelve firms participate in 1,180 reported relationships, accounting for 13.5\% of the national total. Co-linked firms are particularly extreme as recurrent winners: each appears in an average of 253 such relationships, compared with just 4.4 for other firms in the cover graphs (Table~\ref{tab:groups}). Most of their contract wins also occur on exposed tenders (72\%, compared with 22\% for other firms). Discaya-linked firms appear more often as recurrent losers, although their share of losses on exposed tenders is close to that of other firms.}

\begin{table}[!ht]
\centering
\caption{\textbf{Reported relationships and tender exposure of Co-linked and Discaya-linked firms.} Values are pooled across all offices and Congresses. `Other firms' are all remaining firms with at least one reported relationship. All values are computed from contested tenders only. A joint venture's bid is credited to each partner. Shaded rows give group averages: depth is averaged over all of the group's relationships, and the other columns over the group's firms. Co-linked and Discaya-linked firms submit substantially more bids and have more reported relationships than other firms. Most of the Co-linked firms' wins are on exposed tenders.}
\small
\resizebox{\linewidth}{!}{%
\begin{tabular}{lrrrrrr}
\hline\noalign{\vskip 3pt}
Firm & \parbox[b]{2.2cm}{\raggedleft Number of bids submitted on contested tenders} & \parbox[b]{2.6cm}{\raggedleft Number of reported relationships as recurrent winner} & \parbox[b]{2.6cm}{\raggedleft Number of reported relationships as recurrent loser} & \parbox[b]{2.0cm}{\raggedleft Average relationship depth} & \parbox[b]{2.4cm}{\raggedleft Percentage of wins that are exposed tenders} & \parbox[b]{2.4cm}{\raggedleft Percentage of losses that are exposed tenders} \\
\hline
\rowcolor{gray!18}\textbf{Co-linked (3 firms)} & 1,247 & 253.0 & 35.7 & 11.1 & 72\% & 62\% \\
\quad Sunwest & 1,490 & 347 & 38 & 13.5 & 90\% & 57\% \\
\quad Hi-Tone Construction & 2,121 & 307 & 66 & 10.6 & 58\% & 64\% \\
\quad FS Co Builders & 130 & 105 & 3 & 4.7 & 68\% & 64\% \\[2pt]
\rowcolor{gray!18}\textbf{Discaya-linked (9 firms)} & 1,411 & 20.9 & 26.1 & 5.9 & 14\% & 33\% \\
\quad Alpha \& Omega & 3,048 & 68 & 75 & 5.5 & 20\% & 38\% \\
\quad St.\ Gerrard Construction & 2,020 & 36 & 35 & 6.3 & 18\% & 32\% \\
\quad St.\ Timothy Construction & 3,177 & 34 & 70 & 6.3 & 18\% & 39\% \\
\quad St.\ Matthew & 1,229 & 23 & 11 & 7.0 & 27\% & 31\% \\
\quad Elite & 462 & 11 & 3 & 3.7 & 9\% & 29\% \\
\quad Amethyst Horizon Builders & 536 & 9 & 10 & 4.5 & 3\% & 36\% \\
\quad Great Pacific Builders & 892 & 4 & 8 & 6.4 & 13\% & 29\% \\
\quad YPR & 799 & 3 & 15 & 5.8 & 16\% & 35\% \\
\quad Way Maker & 533 & 0 & 8 & 8.1 & 0\% & 30\% \\[2pt]
\rowcolor{gray!18}\textbf{Other firms (1,897 unique firms)} & 240 & 4.4 & 4.4 & 9.0 & 22\% & 30\% \\
\hline
\end{tabular}%
}
\normalsize
\label{tab:groups}
\end{table}

\begin{figure}[!ht]
    \centering
    \includegraphics[width=\linewidth]{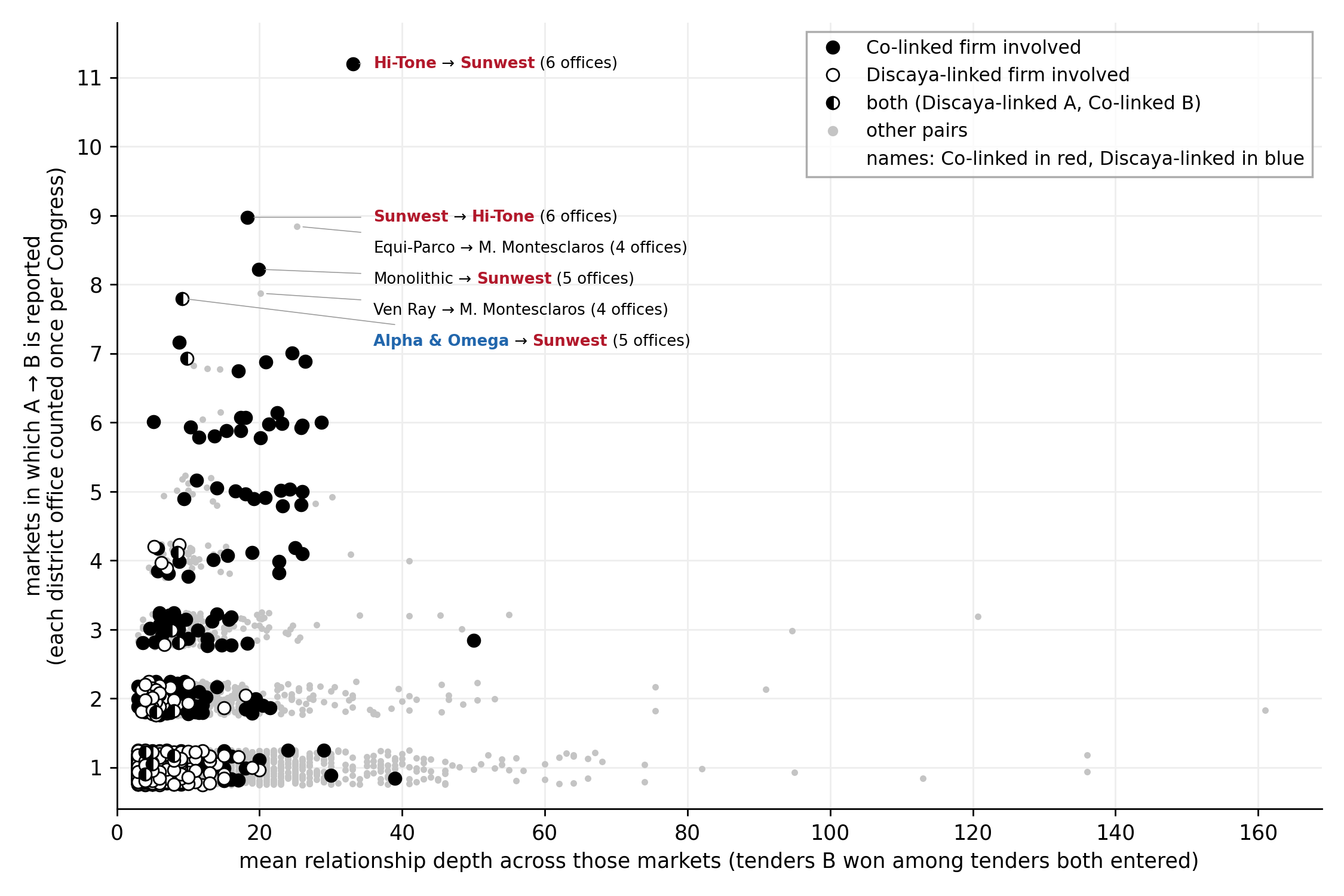}
    \caption{\textbf{Reported pairs across markets.} Each point is
    one of the 7,083 pairs of bidders A~$\to$~B reported in at least one market,
    with A the recurrent loser and B the recurrent winner. A joint venture is
    treated as its own bidder. The vertical axis counts the markets in which the
    pair is reported, with each district office counted once per Congress; the
    horizontal axis gives the average relationship depth across those markets.
    Filled markers show pairs involving a Co-linked firm, open markers pairs
    involving a Discaya-linked firm, and half-filled markers pairs with a
    Discaya-linked recurrent loser and a Co-linked recurrent winner.
    Pairs reported in eight or more markets are labeled with the number of
    district offices in which they are reported; all six are reported in each
    of the three Congresses. Points are spread vertically to reduce overlap.}
    \label{fig:pair_markets}
\end{figure}

{Co-linked firms also stand out because of how often their relationships reappear across markets (Fig~\ref{fig:pair_markets}). Most directed bidder pairs are reported in only one market, but 25 of the 32 pairs reported in six or more markets involve a Co-linked firm. The two most widely recurring directed pairs are Hi-Tone $\rightarrow$ Sunwest and Sunwest $\rightarrow$ Hi-Tone. Their relative winning records vary across markets, from sharing wins in one DEO to one firm overwhelmingly dominating the other. In Romblon, for example, Hi-Tone repeatedly lost to Sunwest across all three Congresses: 28 of 36 shared tenders in the 17th Congress, 72 of 74 in the 18th, and all 50 in the 19th. The relationships between Co-linked and Discaya-linked firms are more one-sided. All 34 reported relationships between the two groups run from Discaya-linked to Co-linked firms. Across 862 shared tenders, Co-linked firms won 373 contracts, while Discaya-linked firms won only six.}

\begin{figure}[!ht]
    \centering
    \includegraphics[width=\linewidth]{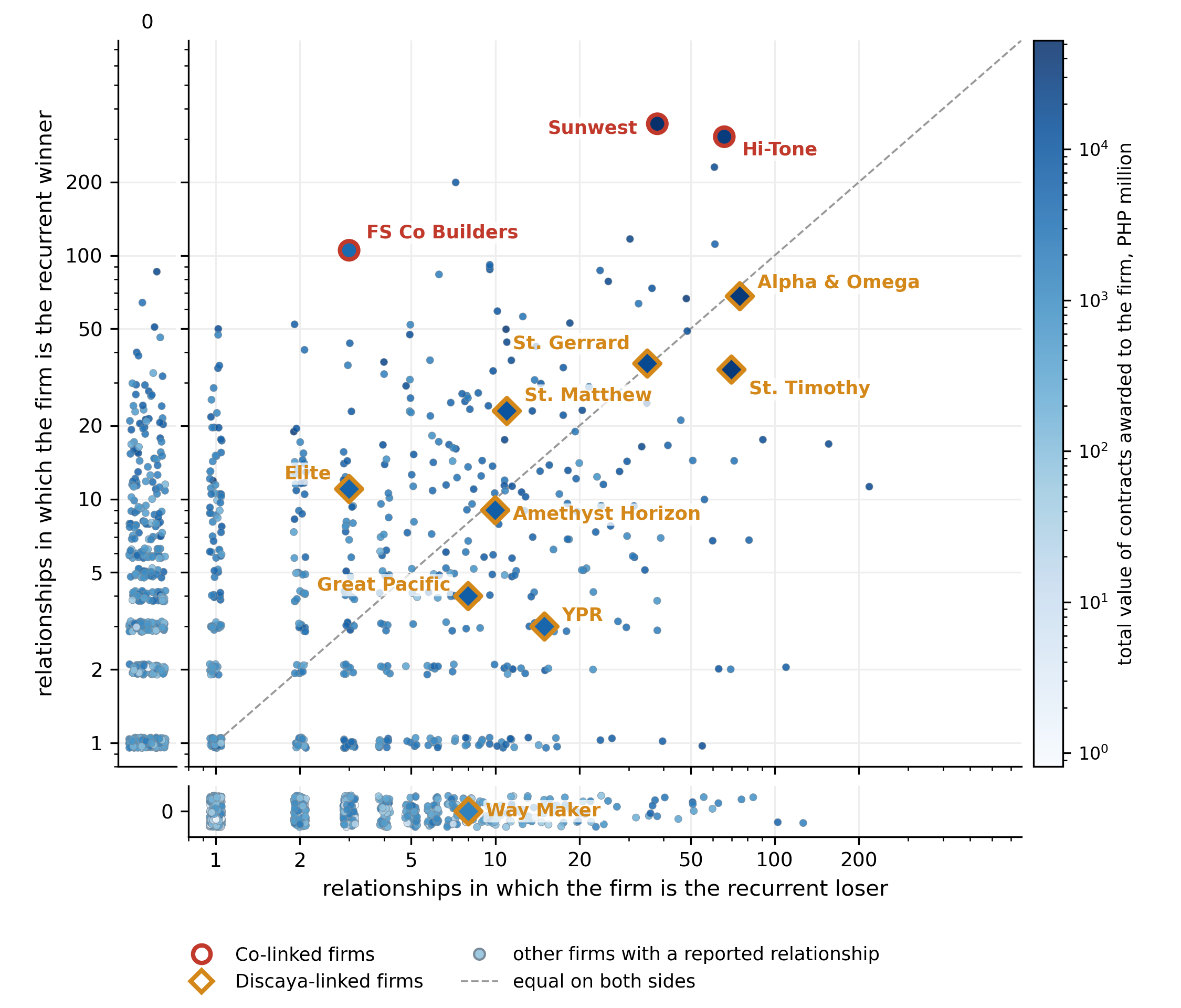}
    \caption{\textbf{Recurrent-winner and recurrent-loser relationships by
    firm.} Each point is one of the 1,909 firms with a reported relationship,
    pooling all offices and Congresses. The horizontal axis counts relationships
    in which the firm is the recurrent loser; the vertical axis counts those in
    which it is the recurrent winner. {Firms in the main panel appear as
    recurrent winners in some markets and recurrent losers in others. Those
    near the diagonal hold the two roles about equally often; those farther
    from it predominantly occupy one.} Co-linked and Discaya-linked firms are outlined and labeled;
    color shows the total value of contracts awarded to each firm. {Both axes
    are logarithmic and start at 1. Firms with no relationship on one side are
    drawn in separate strips: firms that are only recurrent winners on the
    left, firms that are only recurrent losers below.}}
    \label{fig:firm_in_out}
\end{figure}

{The changing roles of Sunwest and Hi-Tone raise a broader question about firms' positions in the cover graph. Persistent differences in competitiveness might produce relatively stable winning--losing hierarchies, with stronger firms repeatedly winning against weaker rivals. Yet some firms occupy both sides of reported relationships within the same geographic area, where differences in local market conditions offer a less convincing explanation for their changing roles. We examine how common these mixed roles are and which firms exhibit them. Fig~\ref{fig:firm_in_out} compares the number of relationships in which each firm appears as a recurrent winner or recurrent loser, pooling across markets. In our sample, 431 firms occupy both roles in different markets within the same region. Sunwest and Hi-Tone are extreme outliers in their recurrent-winner relationships, but also appear as recurrent losers in 38 and 66 relationships, respectively. Almost all of these flagged relationships occur in Bicol, where both firms are major players. The Discaya-linked firms have more varied roles: some appear about equally often as recurrent winners and losers, while others predominantly occupy one position.}

\begin{figure}[!ht]
    \centering
    \includegraphics[width=\linewidth]{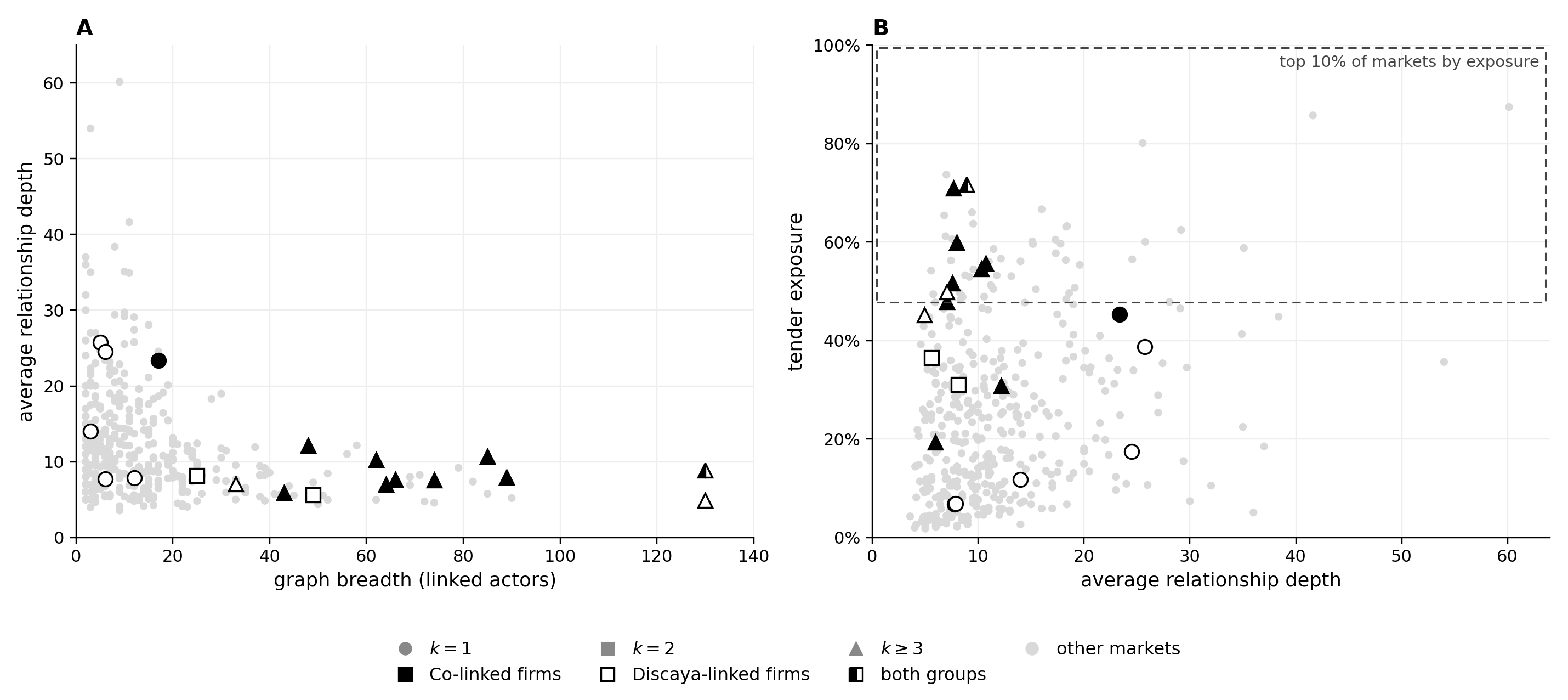}
    \caption{\textbf{Cover-graph characteristics of the markets with the highest Co-linked and Discaya-linked bidding participation.} The comparisons in Figs~\ref{fig:breadth_depth} and~\ref{fig:depth_exposure} are reproduced, highlighting for each group the ten markets (an office in one Congress) in which its firms bid on the largest share of all tenders. Filled markers indicate Co-linked markets, open markers Discaya-linked markets, and the half-filled marker a market appearing in both groups; marker shapes show the maximum $k$-core class, and gray points are all other markets. Co-linked markets have a mean tender exposure of 51\%, twice the average of 25\% across all markets with a cover graph; nine have maximum $k\geq3$, and they include the market with the highest maximum $k$ and the broadest cover graph in the data (Albay 2nd, 19th Congress). Discaya-linked markets span all three classes, with a mean exposure of 32\%. \textbf{A}, graph breadth and average relationship depth. \textbf{B}, average relationship depth and tender exposure; the dashed box marks tender exposure above 48\%, close to the top 10\% of markets within each Congress (above 45--50\%).}
    \label{fig:named_markets}
\end{figure}

{Finally, we examine how the offices where these firms participate most heavily compare with others in the sample. Fig~\ref{fig:named_markets} reproduces the comparisons in Figs~\ref{fig:breadth_depth} and~\ref{fig:depth_exposure}, highlighting the ten markets with the highest bidding participation by each group. In these, Co-linked firms submitted bids on 32--72\% of all tenders, compared with 36--52\% for Discaya-linked firms. The markets with high Discaya-linked participation are spread across all three structural classes and vary widely in relationship depth and tender exposure. On the other hand, the markets with high Co-linked participation include some of the broadest and most interconnected cover graphs in the sample, several with exceptionally high tender exposure.}

\clearpage
\section{Discussion}

The cover networks reveal different patterns in the allocation of winning opportunities and supporting roles, consistent with possible forms of cover bidding. In simpler configurations ($k=1$), recurrent losers are usually connected to a single recurrent winner. This resembles arrangements in which each designated winner relies on a separate set of supporting bidders. More interconnected configurations ($k\geq3$) involve several recurrent winners linked through overlapping groups of losers, resembling arrangements in which supporting bidders provide cover for multiple designated winners \cite{conleydecarolis2016}. Greater interconnectedness does not necessarily imply deeper relationships. The most extreme combinations of relationship depth and tender exposure occur in intermediate ($k=2$) networks (see Fig~\ref{fig:depth_exposure}), where a relatively small number of relationships can extend across most of an office's contested tenders. These differences may reflect competing demands on the internal and external stability of collusive arrangements. Smaller groups may be easier to coordinate \cite{levensteinsuslow2006}, but more vulnerable to independent competitors. Involving more firms may reduce that vulnerability while making cooperation harder to sustain \cite{harrington2023}. Intermediate configurations could represent a balance between these competing pressures.

Sustaining cooperation requires participants to receive sufficient benefits from an arrangement. In bidding rings without side payments, these benefits must come through winning opportunities \cite{mcafeemcmillan1992,pesendorfer2000}. Firms occupying both recurrent-winning and recurrent-losing positions could therefore be exchanging support across tenders. Ishii's study of Japanese procurement identifies a mechanism for such exchanges: firms that had supported particular competitors through earlier losing bids were subsequently more likely to win when those competitors participated \cite{ishii2009}. S6 Fig compares the contract values won and lost along reported relationships. For some firms, the totals are closely matched, consistent with the possibility that earlier support is repaid through subsequent awards. Others exhibit substantial imbalances, which need not rule out reciprocal arrangements if winning opportunities are allocated according to firms' size or market position. Both patterns warrant closer examination where firms occupy both roles. Even more suspicious may be the pairs of firms that occupy both roles against each other. Of the 29 pairs reported in both directions (i.e., $A\to B$ and $B\to A$) within a market, four involve two Co-linked firms and one involves two Discaya-linked firms. Six others involve firms that also bid together as joint ventures. In Quezon 4th during the 18th Congress, for example, two firms entered 145 tenders together, won 67 and 64 of them, and also bid jointly on two contracts in the same office (S3 Fig).

Other arrangements may sustain participation without distributing winning opportunities among all members. In highly interconnected networks, loser-only actors (those that appear only as recurrent losers in their market) rarely win contracts within their focal offices, and they win less often elsewhere than loser-only actors in simpler networks (Table~\ref{tab:loser_winrates}). If these firms participate in a coordinated arrangement, their continued involvement may depend on payments or subcontracts rather than future awards. Common ownership could also align the interests of legally distinct firms without requiring compensation between them. Testimony by a former DPWH undersecretary, given as a state witness, describes payments known as \emph{sahod} (Filipino for ``salary'') to losing contractors \cite{mendoza2026sahod}, providing an example of compensation outside contract awards. The same witness nevertheless acknowledged that ``sometimes, there are legitimate bidders'' \cite{mendoza2026sahod}. Some recurrent losers identified by our screen may therefore be independent competitors whose bids repeatedly fail against stronger firms or designated winners.

The possible involvement of procurement authorities extends the question of coordination beyond the bidding firms. Officials can influence which contractors compete, how contracts are allocated, and whether agreed allocations are maintained \cite{tanaka2016,lambertsonin2006}. Allegations arising from the DPWH investigations describe official involvement in project allocation and contract awards \cite{mendoza2026sahod}, illustrating how bid rigging and corruption may reinforce one another \cite{hudongarzon2016}. Our analysis of contract budgets points to another potentially important stage, even before bidding begins: project preparation and approval. Contracts in more interconnected markets tend to have a greater share budgeted just below the delegated approval ceilings. After the PHP 150 million ceiling was introduced, budgets just below it were about three times as common among exposed tenders as in markets without reported relationships (S1 Table). This concentration just below the DEO-level ceilings is consistent with incentives to structure projects around administrative limits, as documented for Czech procurement thresholds \cite{palgutapertold2017}. Decisions around project preparation and budget approval, which involve the legislative and executive branches \cite{dbm2016budgetcycle}, therefore also warrant examination.

{Tender exposure measures how far reported relationships extend across an office's procurement activity, complementing the organizational information provided by network structure. Its upper tail identifies markets where reported relationships are particularly pervasive and may warrant closer scrutiny, especially where high exposure recurs across Congresses (as in the cases of Isabela 1st, Quirino, and Albay 1st). Highly exposed offices are found in all three structural classes and across Luzon, Visayas, and Mindanao (Fig~\ref{fig:exposure_map}). For investigative screening, we propose prioritizing offices in the highest decile of tender exposure within each Congress. In the 19th Congress, this corresponds to an exposure cutoff of approximately 50\%. The 20 offices meeting this threshold are named in Fig~\ref{fig:exposure_map}. These include Bulacan 1st and Davao Occidental, two of the three DEOs referred for suspected bid rigging.}

{The method identifies pairs in which one bidder wins unusually often against another, not collusive agreements. Its null model assigns equal winning probabilities to bidders within each tender and therefore cannot account for persistent differences in costs, capacity, specialization, or eligibility. Nevertheless, the model provides a simple reference that is sensitive to the pattern of interest: one firm winning repeatedly against the same rival. It requires neither estimates of firms' costs nor assumptions about their competitive strength. Estimating differences in firm competitiveness from historical winning records introduces its own problem: those records may themselves reflect coordination or preferential treatment, so that the estimates absorb some of the patterns the screen is intended to detect. The equal-probability null model is transparent and tractable: each p-value can be computed exactly.}

{Our approach focuses on cover bidding and requires repeated encounters within the observation period. It may therefore miss other forms of bid rigging, such as bid suppression, in which firms refrain from bidding against one another. For instance, single bidding, which suppression can produce, is about twice as common in markets without a reported relationship (31.9\% of contracts) as in markets with one (15.1\%). The screen may also miss bid rotation when each participating firm's winning frequency is close to that predicted by the null model. Markets with fewer tenders offer fewer opportunities for repeated encounters, so low exposure in these markets is weaker evidence of competitive bidding. Establishing whether a reported relationship reflects collusion requires additional evidence on bids, ownership, contractor capabilities, and procurement decisions.}

\section{Conclusion}

This study develops a network-based method for identifying unusual winning--losing relationships in public procurement using bidder and winner identities. Applied to Philippine public works contracts, the resulting directed networks reveal distinct patterns in the distribution of winning and supporting roles. Simpler networks tend to concentrate recurrent losses around particular winners, while more interconnected networks involve overlapping relationships among multiple winners and persistent losers. Some of the deepest relationships occur in intermediate networks, illustrating that greater interconnectedness does not necessarily imply more persistent bidding relationships.

We introduce tender exposure to measure how extensively reported relationships occur within procurement markets. High exposure can arise from a few deep relationships or many shallower ones, allowing the indicator to identify markets warranting closer scrutiny regardless of network structure. Although the method cannot establish collusion, and sees only arrangements that leave a trail of repeated losses, it indicates high collusion risk and provides an interpretable basis for examining possible cover-bidding organization and prioritizing further investigation using routinely collected procurement records.

\section*{Data availability}
The contract-award and bidding records are publicly available from the DPWH Transparency Portal (\url{https://transparency.dpwh.gov.ph/}). The extract analyzed in this study, together with all tested pairs and their test results, will be made available on Zenodo upon publication (\url{https://doi.org/10.5281/zenodo.24264898}).

\section*{Acknowledgments}
JK acknowledges support from the HORIZON-CL2-2024-DEMOCRACY-01 under grant agreement No 101177405.

\section*{Use of AI tools}
IV used generative AI tools (Claude, Anthropic; ChatGPT, OpenAI) to help write code for analysis and visualizations, locate and check sources, and revise the text. IV tested the code, verified the results against the data, and checked the cited sources. The authors made all analytical and interpretive decisions and take full responsibility for the content of the article.

\bibliography{plos_bibtex}

\clearpage
\section{Supporting information}

\paragraph{S1 Fig. Probability of bidding at an away office by distance and administrative-region boundary.}
The figure compares bidding at offices inside and outside the firm's home region over broad distance bands. A firm's home office is the office where it bid most often in that Congress. All other offices are away offices. The home region is the administrative region of the firm's home office. Distances are between office centroids (median project coordinates).

\paragraph{S2 Fig. Abra DEO, 17th to 19th Congress.}
Abra is the market containing the two deepest relationships in the data. Arrows point from recurrent loser to recurrent winner; labels give the tenders the loser lost to the winner over the tenders both entered. Letters are anonymous and identify the same actor in every panel. In the 19th Congress, actor L is the recurrent winner in seven relationships, including 241 losses by J in 376 shared tenders, and the recurrent loser in a relationship with Q (172 of 198). The maximum core is $k=1$ in the 17th Congress and $k=2$ in the 18th and 19th; exposure rises from 27\% to 35\% to 87\%.

\paragraph{S3 Fig. Quezon 4th DEO, 17th to 19th Congress.}
Arrows point from recurrent loser to recurrent winner; labels give the tenders the loser lost to the winner over the tenders both entered. Letters are anonymous and identify the same actor in every panel. In the 18th Congress, actors B and P co-bid on 145 tenders. B won 67 and P won 64, so both directions were reported.

\paragraph{S4 Fig. Actors in the cover graph against actors in a tested pair.}
Each point is one of the 407 markets with a reported relationship. Lines mark 10\%, 30\%, and 100\% of the actors in a tested pair being in the cover graph.

\paragraph{S5 Fig. Contract budgets by exposure status.}
Curves compare exposed tenders, other contested tenders in markets with a cover graph, and contested tenders in markets with no cover graph, by Congress. Vertical dotted lines mark the approval-ceiling amounts. The first PHP 5 million bin exceeds the displayed range; its share is given in the inset. Summary statistics are in S1 Table.

\paragraph{S6 Fig. Contract value won and lost on each firm's relationships.}
For each firm with a reported relationship, the value of contracts won on relationships in which it is the recurrent winner against the value of contracts lost on relationships in which it is the recurrent loser, all offices and Congresses combined. {Firms with no value on one side are drawn in separate strips: firms that only won on the left, firms that only lost below.}

\paragraph{S1 Table. Approved budgets of contested contracts, by Congress, office class, and exposure.}
Offices, contracts, mean approved budget, share at or below PHP 5 million, and shares within 5\% below each ceiling, by Congress, (A) by the office's maximum $k$-core and (B) by exposure.

\paragraph{S2 Table. Budgets just below the approval ceilings and at round amounts, by fiscal year.}
Contested contracts budgeted within 5\% below each approval ceiling (PHP 50, 100, and 150 million) and below round amounts that were never ceilings, by fiscal year.

\paragraph{S1 File. Cover graphs of all markets with a reported relationship.}
{The 22-page PDF draws the cover graphs of all 407 DEO $\times$ Congress markets with at least one reported relationship under the null model, ordered within each Congress by maximum $k$-core and then by exposure. Panels use common node-role colors and common edge-width scaling and report maximum $k$-core, exposure, relationships, graph breadth, and maximum relationship depth.}

\end{document}